\documentclass[prd, twocolumn, nofootinbib, floatfix]{revtex4-1}

\usepackage{amsmath}
\usepackage{graphicx}
\usepackage{dcolumn}
\usepackage{bm}
\usepackage{epsfig}
\usepackage{epstopdf}
\usepackage{amssymb,latexsym,mathrsfs}
\usepackage{graphicx}
\usepackage{color}
\usepackage{hyperref}

\hypersetup{
    colorlinks=true,
    linkcolor=red,
    citecolor=blue,
} 

\newcommand{\be}{\begin{equation}}
\newcommand{\ee}{\end{equation}}
\newcommand{\bs}{\begin{split}} 
\newcommand{\bea}{\begin{eqnarray}}
\newcommand{\eea}{\end{eqnarray}}

\newcommand{\om}{\Omega_m}

\newcommand{\Oe}{\Omega_{e}} 

\newcommand{\h}{{\cal \bar{H}}}
\newcommand{\sigm}{\sigma_{\ln M_0}} 
\newcommand{\lnm}{\ln M_{{\rm bias},0}}

\begin{document}

\title{Cluster Probes of Dark Energy Clustering} 
\author{Stephen A.\ Appleby$^1$, Eric V.\ Linder$^{1,2}$, Jochen Weller$^{3,4,5}$} 
\affiliation{$^1$ Institute for the Early Universe WCU, Ewha Womans 
University, Seoul 120-750 Korea} 
\affiliation{$^2$ Berkeley Lab \& University of California, Berkeley, 
CA 94720, USA} 
\affiliation{$^3$ Universit\"{a}ts-Sternwarte M\"{u}nchen, Ludwig-Maximilians Universit\"{a}t M\"{u}nchen, Scheinerstr.~1, 
D-81679, M\"{u}nchen, Germany} 
\affiliation{$^4$ Excellence Cluster Universe, Technical Univserity Munich, Boltzmannstr.~2, D-85748 Garching, Germany} 
\affiliation{$^5$ Max-Planck-Institut f\"{u}r Extraterrestrische Physik, Giessenbachstr., D-85748 Garching, Germany}

\begin{abstract}
Cluster abundances are oddly insensitive to canonical early dark energy.  
Early dark energy with 
sound speed equal to the speed of light cannot be distinguished from 
a quintessence model with the equivalent expansion history for $z<2$ but 
negligible early dark energy density, despite the different early 
growth rate.  
However, cold early dark energy, with a sound speed much smaller than the 
speed of light, can give a detectable signature.  Combining cluster 
abundances with cosmic microwave background power spectra can determine 
the early dark energy fraction to 0.3\% and distinguish a true sound speed of 
0.1 from 1 at 99\% confidence.  
We project constraints on early dark energy from the Euclid cluster survey, 
as well as the Dark Energy Survey, using both current and projected Planck 
CMB data, and assess the impact of 
cluster mass systematics.  We also quantify the importance of dark energy 
perturbations, and the role of sound speed during a crossing of $w=-1$. 
\end{abstract}

\date{\today} 

\maketitle

\section{Introduction} 

Understanding the physics behind cosmic acceleration requires clear 
characterization of the properties of dark energy.  This includes its 
dynamics, e.g.\ equation of state behavior $w(z)$, its degrees of freedom, 
e.g.\ perturbations or sound speed $c_s(z)$, and its persistence, i.e.\ 
presence of dark energy at high redshift.  Distance measurements are 
most sensitive to the first of these properties \cite{suzuki,wigglez,boss}.  
Cosmic microwave background (CMB) data can constrain the second and third to 
some extent \cite{2003MNRAS.346..987W,2004PhRvD..69h3503B,09010916,dhl,10105612,11034132,11105328,11060299} in particular if correlated with large scale structure observations.  
Growth of large scale structure 
might be expected to also be affected by perturbations and early dark energy 
\cite{doranrob,Wetterich04,fll1,grossi,fll2,dhl,alam}, but normalization to 
the present growth amplitude or abundance removes almost all this 
sensitivity \cite{fll1}. 

For perturbations in the dark energy to have an appreciable influence on 
the matter power spectrum and large scale clustering, two conditions are 
necessary.  The first is that the dark energy equation of state, or 
pressure to density, ratio $w$ must be significantly different from $-1$, 
the cosmological constant value, since the influence of perturbation enters 
with a prefactor $1+w$ (see \cite{dhl} for analytic scalings).  Since 
data constraints indicate that at low redshift $w\approx-1$, we need 
persistence of dark energy to high redshift, where $w$ can be significantly 
different from $-1$, approaching $w\approx0$.  Thus we talk about early 
dark energy, that may have a fraction of the critical density 
$\Omega_{de}(z_{lss})\approx 10^{-2}$ at the CMB 
last scattering surface rather than $\Omega_\Lambda(z_{lss})\approx 10^{-9}$.  

The second requirement is that the sound horizon $c_s H^{-1}$ of the dark 
energy perturbations be well within the Hubble scale $H$.  Dark energy 
clusters only on scales outside its sound horizon and smaller than the 
Hubble scale, $H<k<H/c_s$, where $k$ is the perturbation wavemode, in 
the same way that matter clumps only on 
scales greater than its own Jeans scale.  Thus, inclusion of dark energy 
perturbations per se is not the key, but perturbations that can grow. 
Therefore we require that the sound speed be small compared to the speed 
of light, $c_s\ll1$.  This is referred to as cold dark energy 
\cite{10072188,10105612}.  In this article we explore signatures of cold 
early dark energy on galaxy cluster abundances. 

In Sec.~\ref{sec:model} we discuss the influence of dark energy in 
terms of its expansion history and perturbations, especially on the matter 
power spectrum and cluster mass function.  We compute the cluster 
abundances in Sec.~\ref{sec:signal} in different models for forthcoming 
surveys to determine the signal to noise of the dark energy signature.  
Including cosmological parameter and observational systematics covariance 
in Sec.~\ref{sec:fisher} we 
project constraints on the dark energy properties.

\section{Dark Energy Effects on Matter Clustering} \label{sec:model} 

Dark energy acts to suppress the growth of matter structures through 
increasing the Hubble friction and reducing the matter source term 
(see, e.g., \cite{linjen} for detailed discussion).  Early dark energy, 
through its persistence to higher redshifts, strengthens the suppression. 
Even if at early times the dark energy has $w\approx0$, i.e.\ acts 
roughly like matter in the expansion, the reduction in the source term for 
matter perturbations causes suppression.  

Of course the dark energy itself 
has perturbations, as any fluid with $w\ne-1$ must, but these generally 
provide negligible contribution to the Poisson term sourcing the matter 
perturbations.  For example, \cite{dhl,10045509} show that for a canonical 
sound speed $c_s=1$, the ratio of the dark energy perturbation power spectrum 
to the matter power spectrum goes as $(k/H)^{-4}$.  Even at wavemode 
$k=0.01\,h$/Mpc, only accessible to very large scale structure surveys, 
dark energy only contributes $\sim 10^{-8}$ as much power as matter does. 
Therefore canonical dark energy, even early dark energy, has negligible 
effect on the matter growth, power spectrum, or cluster abundances other 
than through its expansion effects on the growth.  If this is normalized 
out by fixing the mass fluctuation amplitude $\sigma_8$ today, then 
there is remarkable insensitivity to the presence of early dark energy. 

This was demonstrated in detail through computing the matter power 
spectrum and halo mass function (HMF) from N-body simulations and showing 
their close agreement with $\Lambda$CDM computations \cite{fll1}.  This was 
also seen for the HMF for a different early dark energy model in 
\cite{grossi}, assuming the $\Lambda$CDM linear collapse threshold 
$\delta_c$ within the spherical collapse formalism, and \cite{fll2} 
derived that $\delta_c$ is indeed near the $\Lambda$CDM value.  The 
insensitivity is robust to the manner of identifying halos and the 
specific mass function used.  At redshifts $z>0$, few-tens of percent 
level deviations can arise in cluster abundances with mass 
$M>10^{14}M_\odot/h$ due to the differing growth histories \cite{fll1}.  
The internal structure of clusters, e.g.\ their concentrations, could also 
show signs of the differing growth history of early dark energy 
\cite{grossi}. 

Differences in the HMF within another early dark energy model were found 
in \cite{alam} but this is due to strongly differing values of the present 
matter fluctuations, i.e.\ $\sigma_8$.  It is not due 
to the inclusion of dark energy perturbations; note that 
\cite{fll1} included perturbations in the initial linear power spectrum 
of the simulations 
and as mentioned above \cite{dhl} calculated the effect of perturbations 
and found them to be negligible for canonical dark energy.  These results 
leave open the possibility that cold early dark energy could give detectable 
effects on the HMF (\cite{dhl} found signatures from it in the matter and CMB 
power spectra).  We investigate this in the next section. In general
we should note that a cold early dark energy component is expected to
be more inhomogenous then a component with a sound speed of
$c_s=1$. On the non-linear level this could lead to stronger
backreaction effects also on the matter distribution and hence
altering the HMF on a level beyond the one included in the change of
the linear matter power spectrum. However in order to simulate this
effect properly we would need to include the scalar field on a grid in
N-body simulations. This is an extremely demanding task, which so far
has only been addressed in the context of scalar-tensor theories
\cite{Oyaizu:08,Li:12,Puchwein:13}. We follow the approach to only
include the changes in the linear power spectrum and propagate to the 
nonlinear regime in the standard way as described
below. We expect that this will result in conservative constraints,
which might be tighter once the full nonlinear behaviour is taken
into account.

\section{Cluster Mass Function Signals} \label{sec:signal} 

The halo mass function is generally calculated from a fitting form that 
is a function of the linear mass fluctuation amplitude on the scale 
corresponding to the cluster mass, $\sigma(M,z)$.  A popular modern form, 
adopted here, is the Tinker et al.~\cite{tinker} mass function, 
\bea 
\frac{dn}{dM}&=&f(\sigma)\,\frac{\bar\rho_m}{M}\frac{d\ln\sigma^{-1}}{dM}\\ 
f(\sigma)&=&A\left[\left(\frac{\sigma}{b}\right)^{-d}+1\right]\, 
e^{-c/\sigma^2}\\ 
\sigma^2(M,z)&=&4\pi\int dk\, k^2\,P(k,z)\,W^2(kR) \ ,
\eea 
where $n$ is the cluster abundance, $f$ is the multiplicity function, 
$P$ is the linear matter power spectrum, and $W$ is the window function on 
scale $R$ corresponding to mass $M$. 

The linear power spectrum for a dark energy model, cold and early or not, 
can be calculated from a modified version of CAMB \cite{camb}.  This 
includes perturbations in the dark energy; see the discussion later of 
their special importance for cold early dark energy. 

The Tinker 
et al.\ mass function has been investigated for universality with
respect to cosmology, however not for early dark energy models.  
As mentioned above, one would need to carry out involved N-body simulations 
of each such model, including dark energy fields to follow their 
perturbations (and testing halo identification with respect to 
virialization).  
On the other hand, \cite{fll1,grossi} showed that older mass functions 
such as Jenkins et al.~\cite{jenkins}, Sheth-Tormen \cite{shethtormen}, 
and Warren et al.~\cite{warren}, while also calibrated from $\Lambda$, 
still agreed as well with N-body simulations of early dark energy 
cosmologies.  Furthermore, \cite{fll2} demonstrated that the linear 
collapse threshold for early dark energy, as enters the Sheth-Tormen 
and similar excursion set approaches, was quite similar to $\Lambda$CDM. 
For low sound speed, the clustering of the dark 
energy fluid makes it act more like standard matter, and hence this is 
even more true as computed by \cite{pace}, and so 
the mass function form should be even more acceptable, although a
final verdict requires the inclusion of an inhomogenous cold early
dark energy component in the simulations.
For conservative reasons we assume the Tinker et al.\ HMF holds for the cold early dark 
energy models, with the same constant coefficients given in \cite{tinker}; 
we note that this HMF is commonly used in the literature for 
non-$\Lambda$CDM cosmologies, although eventually simulations should 
verify the level of universality. 

The signal to noise of a cold early dark energy component in the cluster 
mass function is given by the deviation of the abundance relative to a 
fiducial model, say, $\Lambda$CDM with the same cosmological parameters 
other than those for dark energy, 
\be 
\frac{S}{N}=\frac{N_{\rm model}-N_{\rm fid}}{\delta N_{\rm model}}\ , 
\ee 
where this can be evaluated for each mass bin and redshift bin (or summed 
in quadrature over them), and $\delta N$ 
represents the measurement uncertainty of the cluster abundance (e.g.\ 
Poisson error) within the bin. 

Survey characteristics enter through the mass threshold, below which 
clusters cannot be detected, redshift range, and the Poisson error that 
follows from the survey volume (as well as systematic uncertainties 
discussed in the next section).  We consider specific surveys in the next 
section.  

\begin{figure}[htbp!] 
\includegraphics[width=\columnwidth]{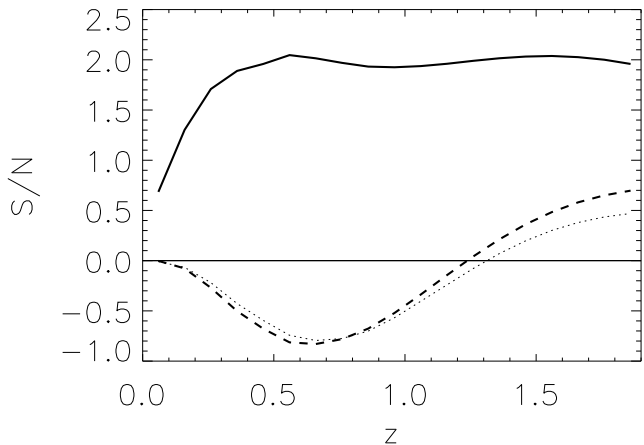} 
\caption{Cluster abundance deviations from $\Lambda$CDM are shown in 
terms of signal to noise.  Early dark 
energy with $c_s=1$ (dashed curve) agrees closely with $\Lambda$CDM 
despite the different growth and expansion history.  Matching the 
EDE expansion history with a quintessence model with no early dark 
energy (dotted curve: $w_0=-0.99$, $w_a=0.04$) shows virtually no 
difference in their cluster abundances.  However, cold early dark 
energy with $c_s=0$ (solid curve) gives slightly more significant differences.  
}
\label{fig:sovern} 
\end{figure}

Figure~\ref{fig:sovern} plots the signal to noise for distinguishing 
three different 
cosmologies relative to $\Lambda$CDM with the Euclid cluster survey 
(discussed further in the next section).  
The early dark energy density is taken to be of the Doran-Robbers form 
\cite{doranrob}, asymptoting to an early time fraction $\Omega_e=0.01$, 
at the current limits when the sound speed is fixed to the speed of light.  
The current equation of state is 
chosen to be $w_0=-0.99$, almost the same as a cosmological constant.

When the sound speed $c_s=1$, then the cluster abundance agrees closely 
with $\Lambda$CDM at each redshift, always with $S/N<1$ in each 0.1 
redshift bin.  Thus, even though their early expansion and growth histories 
are distinct, the cluster abundance is insensitive to these, as found 
in \cite{fll1,grossi}.  To focus on the different growth history, we 
can use the fitting form of \cite{linderrob} that identified nearly 
identical expansion histories between an early dark energy cosmology 
and a corresponding quintessence model that has no early dark energy.  
The rule of thumb is that a quintessence model with the same current 
equation of state $w_0$ and a time variation $w_a\approx5\Omega_e$ 
would have nearly the same expansion history as the early dark energy 
model for $z<2$ (where the cluster observations are).  We find that 
$w_a=0.04$ matches the $\Omega_e=0.01$ early dark energy case to 0.1\% 
in distance and 0.4\% in volume element for $z<2$.  In Fig.~\ref{fig:sovern} 
we see that these matched models have nearly identical cluster abundances, 
despite very different early expansion history (e.g.\ 
$\Omega_{de}=6\times 10^{-9}$ vs 0.01 at $z=1090$). 

However, the cold early dark energy model has significantly different 
cluster abundances.  This simple signal to noise estimation, using the 
number of clusters expected from the Euclid satellite \cite{euclid}, 
shows an easily detectable signal deviating from $\Lambda$CDM, with a 
single bin $S/N\approx2$ for each bin in $0.2\le z\le1.9$.  This appears 
promising.  

Note that for cold dark energy the treatment of perturbations in the 
dark component is crucial.  Appendix~\ref{sec:apxpert} assesses the 
impact of the perturbations, comparing the standard differential equation 
for the growth factor to full solutions of the Boltzmann equations.  
Appendix~\ref{sec:apxcross} discusses general properties of perturbations 
when the equation of state ratio $w$ crosses $-1$ (although our early 
dark energy model stays at $w>-1$), especially the role of sound speed. 

The total detectability of deviations from $\Lambda$CDM 
will be enhanced by summing over all redshifts, and tomography in mass 
bins helps as well.  Conversely, covariance with other cosmological 
parameters and systematic contributions to the uncertainty above the 
Poisson level will decrease the possible signal.  We incorporate these 
into a more realistic calculation of early dark energy characterization 
in the next section.

\section{Detection of Cold Early Dark Energy} \label{sec:fisher}

\subsection{Survey Characteristics} 

Here we consider prospective parameter constraints arising from 
future Dark Energy Survey (DES \cite{des}) and Euclid clustering data.  
To accurately constrain cosmological parameters, we must take into account systematics that are used to model the various uncertainties associated with the survey selection function and cluster scaling relations (see, e.g., 
\cite{Cunha:2009dz,Cunha:2009rx}).

To calculate the mass of a cluster we assume an approach based on photometric 
richness (additional information, not used here, may come from weak 
gravitational lensing within Euclid, or Sunyaev-Zel'dovich and X-ray 
surveys).   Systematics in 
this relation are treated through a bias and a scatter.  

To account for possible bias between the cluster richness mass estimate 
and the true cluster mass, we define the mass bias as 
\begin{equation}\label{eq:sa1}  
\ln(M_{\rm bias}) = A + B \ln(1+z) \ , 
\end{equation} 
where $A,B$ are nuisance parameters to be marginalised over. We have
accounted for a possible power law evolution with redshift, as
extolled in \cite{Lima:2005tt}. We set the fiducial values as $A=B=0$
and assign the quantities Gaussian priors $\sigma_{\rm A} =
\sigma_{\rm B} = 0.25$. For the theoretical exercise here we neglect
photo-z errors, effects of purity and completeness of the sample, the
covariance between the clustering of clusters and the counts. However
we include the sample variance due to large scale structure and use
the clustering of clusters as additional probe.

We also model the intrinsic scatter around the selection function as a 
lognormal distribution with dispersion
\begin{equation} \label{eq:sa2} 
\sigma_{\ln M} = \sigma_{\ln M_{0}} - 1 + (1+z)^{2\beta} \ . 
\end{equation} 
In \cite{Rykoff:2011xi} it is estimated that $\sigma_{\ln M_{0}} = 0.2$ and we take $\beta = 0.125$. For our forecast analysis we adopt 
conservative Gaussian priors $\sigma(\sigma_{\ln M_{0}})=0.1$, 
$\sigma(\beta)=0.1$.  Given the uncertainty in forecasting the performance 
of experiments, we take the same priors for both DES and Euclid, though over 
different redshift ranges.  In general we find that the survey results are 
not strongly influenced by the priors.  

For the forecast for Euclid, we take the survey mass threshold sensitivity 
limit to be a weakly redshift dependent function varying between 
$M_{\rm lim} \sim 10^{13.5} {h^{-1}} M_{\odot}$ at $z=0.2$ to a roughly constant value of $M_{\lim} \sim 10^{14.1}{ h^{-1}} M_{\odot}$ at
$0.4<z<2$ \cite{euclid,Biviano}, which corresponds to a $3\sigma$ detection limit, assuming a simple overdensity detection threshold.
We bin the clusters in redshift, with bins of width $\Delta z = 0.1$ between 
$z=(0.2,2)$, and in mass, using four bins between $M_{\rm lim}(z)$ and 
$M = 10^{15} {h}^{-1}M_{\odot}$ and one between $M = 10^{15} {h}^{-1}M_{\odot}$ and an 
arbitrarily large upper bound, taken to be $M = 10^{17} {h}^{-1}M_{\odot}$. We have checked that using ten bins 
between $M_{\rm lim}(z)$ and 
$M = 10^{15} {h}^{-1}M_{\odot}$ yields very similar results. 
For the DES analysis, we use a constant mass limit 
$M_{\rm lim} = 1.2 \times 10^{14} {h}^{-1} M_{\odot}$ over the range 
$z=(0,1)$ \cite{Rozo}, and the same number of mass bins as for Euclid.  Survey characteristics used are summarized in 
Table~\ref{tab:survey}.

\begin{table}[!htb]
\begin{tabular}{l|ccc} 
Survey \ & \ Area (deg$^2$) \ & $z$ & \ $M_{\rm lim} (h^{-1}M_\odot)$ \  \\
\hline 
Euclid & 15,000 & 0.2--2 & $\sim 10^{14}$  \\ 
DES & 5,000 & 0--1 & $1.2 \times 10^{14}$  \\ 
\end{tabular}
\caption{Survey characteristics adopted for our forecasts are 
shown for two forthcoming optical cluster surveys, including sky area, 
redshift range $z$, and limiting mass threshold $M_{\rm lim}$. 
}
\label{tab:survey}
\end{table}

\subsection{Cosmological Constraints} 

To explore cluster abundance constraints on cold early dark energy 
cosmology we carry out a Fisher information analysis as a first indication 
of detectability.  In addition to the standard six parameters of 
physical baryon density $\Omega_b h^2$, total matter density $\om$, 
reduced Hubble constant $h$, mass fluctuation amplitude $\sigma_8$, 
scalar perturbation tilt $n_s$, and optical depth $\tau$, we include 
the cold early dark energy parameters of the early dark energy density 
$\Oe$, present equation of state ratio $w_0$, and sound speed $c_s$. 
We take Planck fiducials \cite{Ade:2013zuv} for the standard 
parameters, plus $\Oe=0.009$ and $w_0=-0.97$, within current constraints.  

Since the magnitude of $c_s$ could range substantially from 1 (or more) 
to very small values, and since it enters the perturbation equations 
as $c_s^2$, we take $\ln c_s^2$ as the sound speed parameter.  To test 
whether clusters could possibly distinguish cold early dark energy, unlike 
other probes, we choose a fiducial of $c_s=0.1$ or $\ln c_s^2=-4.6$.  For 
larger $c_s$, perturbations will be suppressed and so the signature will 
be absent, while for smaller $c_s$ the lack of suppression saturates 
\cite{dhl} and so sensitivity to the exact value of $c_s$ degrades.  
Similar properties hold for cosmic microwave background power spectra, 
as seen in Fig.~4 top panel of \cite{10105612}.  
Thus a fiducial $\ln c_s^2=-4.6$ provides good leverage for 
distinguishing cold early dark energy.  We discuss the influence of the 
fiducial value, and a linear rather than log distribution, 
later in this section. 

Taking the survey characteristics for Euclid given above, in conjunction 
with the estimated final Planck cosmic microwave background sensitivity 
\cite{planck}, 
we project constraints on the cosmological parameters.  Note that without 
CMB information, the degeneracies present in the cluster mass 
function prevent meaningful constraints.  For example, the sound speed 
has correlation coefficients exceeding 0.8 amplitude with 
$\Omega_c h^2$ and $n_s$.  Thus, even though we find that 
clusters give stronger unmarginalized estimation of $c_s$ than from CMB, 
for the marginalized uncertainty the CMB leverage dominates. 
Conversely, the CMB has strong degeneracies for the early dark energy 
model, such as on $w_0$, that cluster data breaks. 

\begin{figure*}[htbp!]
\includegraphics[width=\textwidth]{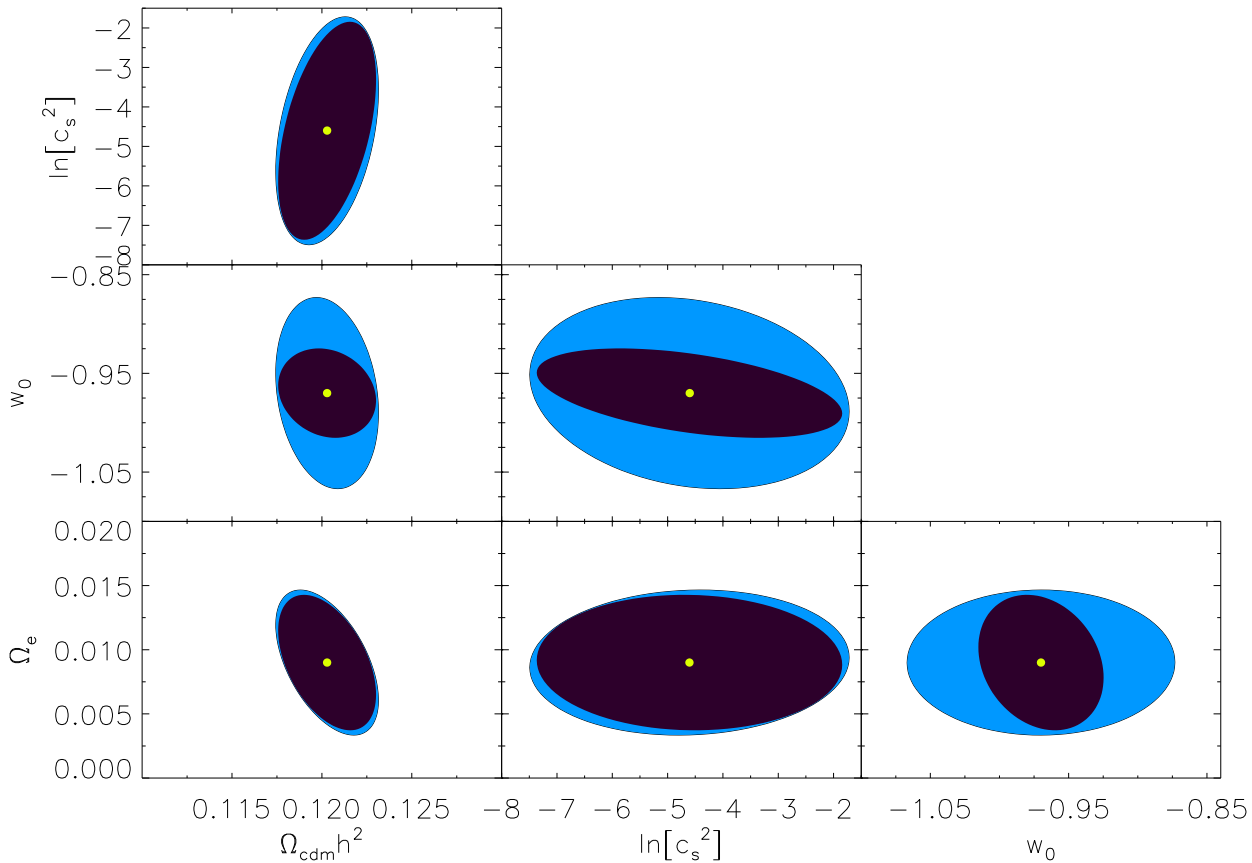}
\caption{68\% confidence level contours from projected Euclid (black) 
and DES (light blue) cluster abundances, with projected Planck CMB data, 
are shown for estimation of pairs of dark cosmology parameters, marginalized 
over all other parameters. The fiducial parameters are indicated by the 
yellow dots.}
\label{fig:triangle}
\end{figure*}

Figure~\ref{fig:triangle} shows the likelihood contours for pairs of dark 
cosmology parameters, marginalized over all other parameters including 
systematics.  The data would distinguish between 0.9\% early 
dark energy density and none, and between actual cold early dark energy (low 
$c_s$) and regular ($c_s=1$, i.e.\ $\ln c_s^2=0$) early dark energy.  
As discussed below, the converse does not hold: if $c_s=1$ in truth, we 
could not distinguish it from cold dark energy since the lack of leverage if 
$c_s\approx1$ means its estimation uncertainty is high. 

Cluster mass systematics do not play a large role.  In fact, the survey 
selfcalibrates $(\sigm,\beta,\lnm,B)$ to uncertainties of 
$(0.033,0.014,0.055,0.057)$.  In particular this means that we do not 
require stringent information on the redshift evolution of the systematics. 
With no priors on the systematics parameters, the early dark energy density 
is determined to $\sigma(\Oe)=0.00352$, for the present dark energy density 
$\sigma(w_0)=0.0309$, and for the sound speed $\sigma(\ln c_s^2)=1.82$.  
The two early dark parameters are only mildly correlated with the standard 
set.  For our fiducial priors on the systematics, 
$(0.1,0.1,0.25,0,25)$, we find 
$\sigma(\Oe)=0.00348$, $\sigma(w_0)=0.0299$, $\sigma(\ln c_s^2)=1.82$. 
Without systematics the uncertainties would be 0.0255, 0.0180, 1.07
respectively.

While a fiducial of cold early dark energy can be distinguished clearly 
from standard early dark energy or no early dark energy, this is not 
true for a fiducial of standard early dark energy.  
Table~\ref{tab:sigmas} gives the marginalized cosmological parameter 
estimation for our baseline fiducial sound speed $c_s=0.1$, 
for fiducial $c_s=1$, and also when sound speed is fixed to $c_s=1$. 
The sound speed $c_s=1$ can be confused with $c_s^2=1/3$ ($\ln c_s^2=-1.1$) 
at $1\sigma$, and we find below that in fact it is 
essentially unconstrained.  This is due to the 
property discussed earlier that near the extremes of the $c_s$ range, not 
only is the exact value of $c_s$ hard to determine, but degeneracies with 
other parameters come into play.  
Similar conclusions have been reached recently on constraints projected 
for weak lensing and galaxy clustering probes \cite{13051942}.  
To test the robustness of this conclusion we also use $c_s^2$ rathan 
than $\ln c_s^2$ as the parameter.

\begin{table}[!htb]
\begin{tabular}{l|ccccc}
Case& $\Omega_{c}h^{2}$ & $\Oe$ &$w_0$&$\ln c_s^2$& $c_s^2$\\ 
\hline 
$c_s^2=0.01$ (fid)$\,$ & $\ $ 0.0018 $\ $ & $\ $0.0035$\ $ & $\ $0.030$\ $ & $1.82$ & $0.018$ \\ 
$c_s^2=1$ (fid)   & 0.0018 & 0.0057 & 0.045  & $1.42$ & $1.43$ \\ 
$c_s^2=1$ fixed & 0.0018 & 0.0047 & 0.042 & -- & -- \\ 
\end{tabular}
\caption{$1\sigma$ constraints from projected Euclid cluster abundance 
and future Planck CMB data (full temperature and polarization) on dark 
energy parameters are compared for different fiducial cases of the sound 
speed.  The last two columns are the errors obtained when the sound 
speed (squared) parameter is taken to be log or linear, respectively. 
}
\label{tab:sigmas}
\end{table}

\begin{table}[!htb]
\begin{tabular}{l|ccccc}
Case& $\Omega_{c}h^{2}$ & $\Oe$ &$w_0$&$\ln c_s^2$& $c_s^2$\\ 
\hline 
$c_s^2=0.01$ (fid)$\,$ & $\ $ 0.0020 $\ $ & $\ $0.0051$\ $ & $\ $0.050$\ $ & $2.78$ & $0.042$ \\ 
$c_s^2=1$ (fid)   & 0.0021 & 0.0045 & 0.053  & $3.58$ & $1.15$ \\ 
$c_s^2=1$ fixed & 0.0021 & 0.0046 & 0.052 & -- & -- \\ 
\end{tabular}
\caption{As Table~\ref{tab:sigmas} but for projected Euclid cluster abundance 
and current Planck CMB temperature data (including WMAP polarisation data). 
}
\label{tab:sigmascurr}
\end{table}

The dark energy sound speed constraint arises predominantly from the CMB 
data.  Since a first Planck data release is available, we can compare 
the constraints using this partial temperature data plus WMAP polarization 
data \cite{wmap9} vs 
the projected future full temperature plus polarization Planck data.  
To use the current Planck data, we modify CosmoMC \cite{cosmomc} and perform a 
Markov Chain Monte Carlo analysis varying over the standard six cosmological 
parameters and in addition the cold early dark energy parameters 
$\Omega_{\rm e}$, $w_0$, and $\ln c_{s}^{2}$.  
These results in combination with projected Euclid cluster data are 
shown in Table~\ref{tab:sigmascurr}.  CMB-only constraints, for current 
data and for projected full Planck data, are shown in Table~\ref{tab:pl_dat}.

\begin{table}[!htb]
\begin{tabular}{l|cccc}
Case& $\Omega_{c}h^{2}$ & $\Oe$ &$w_0$&$\ln c_s^2$\\ 
\hline 
Current (T + WMAP pol)&0.0030&0.0054& 0.350 & 3.72 \\ 
Full (TT,TE,EE)$\,$ & $\ $0.0020$\ $ & $\ $0.0039$\ $ & $\ $0.420$\ $ & 
$\ $1.94$\ $\\ 
\end{tabular}
\caption{Comparison of $1\sigma$ constraints from the CMB-only 
analysis is shown using the current actual data release (temperature 
only Planck with WMAP polarisation) vs the projected full Planck 
temperature and polarization data.  In the MCMC analysis using current 
data the sound speed is only weakly constrained. 
(Note there is a large difference between the $w_{0}$ best fit value for 
the MCMC and the fiducial in the projected Fisher analysis).} 
\label{tab:pl_dat}
\end{table}

\subsection{Survey Comparison} \label{sec:des} 

Cluster abundance data will become available from DES in the near term, 
with Euclid data following several years later.  Euclid will have a larger 
sky area, enhanced redshift range, and somewhat lower mass threshold.  These 
will also impact the systematics control.  
Figure~\ref{fig:triangle} compares the constraints that will be enabled by 
these cluster surveys when combined with Planck data. 

Cluster abundances play an important role in estimating the dark energy 
equation of state today $w_0$, to which the CMB is insensitive, and we 
find that Euclid will provide significantly improved constraints, by a 
factor of $\sim2.5$ better even than DES.  The early dark energy sound speed, 
and energy density, are substantially determined by the CMB data alone.  

For cold early dark energy, its density can be determined to 0.3\% of the 
critical density when combining clusters plus CMB, even fitting simultaneously 
for the sound speed.  Cold early dark energy can clearly be distinguished 
from the canonical case with $c_s=1$, if it really is cold, but if it is 
canonical then cold dark energy cannot be ruled out.

\section{Conclusions} \label{sec:concl} 

The halo mass function describing the abundance of massive galaxy clusters 
is sensitive to properties of both the background expansion and the growth 
of structure.  Since growth involves the expansion history at all earlier 
times, one might hope to use growth as a probe of the presence of early 
dark energy, such as appears in many high energy physics theories.  However, 
cluster abundances and other growth measurements have been previously 
found to have difficulty discriminating early dark energy due to the ability 
of time varying dark energy to mimic its effects.  Here we show 
that cold early dark energy, where the dark energy perturbation effects 
are enhanced, can be distinguished by a combination of cluster abundance 
and CMB data. 

The Euclid satellite cluster survey, in conjunction with CMB data, 
will be able to detect the existence 
of early dark energy with 0.9\% density at the 99\% confidence level, and 
moreover detect that the early dark energy is cold ($c_s\lesssim0.1$) 
rather than canonical ($c_s=1$) at 99\% confidence (if it really is cold). 
Note that early dark energy models obviate the need for the dark energy 
equation of state today to be significantly different from $w\approx-1$ 
in order for sound speed to have a reasonable impact.  Moreover, many 
high energy physics models, such as Dirac-Born-Infeld dark energy or 
various string-inspired models, have specifically cold early dark energy. 

Near term cluster surveys such as DES will have sufficient leverage to 
break degeneracies in CMB data and in combination achieve similar limits 
on $\Oe$ and $c_s$.  The constraints on the present dark energy equation of 
state will have uncertainties $\sigma(w_0)\sim0.06$, improving to $\sim 0.03$ 
with Euclid. 

Interestingly, the complementarity of cluster and CMB data leads to 
good selfcalibration of the cluster mass systematics, including allowance 
for redshift evolution in the scatter and bias.  The Euclid cluster survey 
has sufficient information to determine the uncertainty in the scatter, 
$\sigma(\sigma(\ln M_0))$ to 0.033, for example.  Further reducing 
systematics, or tightening the estimation of $\Omega_c h^2$, would help 
better determine the sound speed, reducing its fractional uncertainty 
by almost a factor of 2. 

These results offer promising signs for the ability of next generation 
cluster abundance measurements to probe the nature of early dark energy. 
Further leverage could come from more precise CMB lensing measurements 
from ground based polarization experiments, and from crosscorrelation 
of the CMB with high redshift tracers of the density field (cf.\ 
\cite{vallinotto}).  While lower 
redshift measurements provide important information on dark energy 
dynamics, higher redshift measurements of structure growth illuminate 
the two other important properties of dark energy: its persistence and 
internal degrees of freedom.

\acknowledgments 

We thank Andrea Biviano, Carlos Cunha, Alireza Hojjati and Eduardo Rozo for helpful discussions.  
SAA is grateful to the Berkeley Center for Cosmological Physics, and 
JW to the Institute for the Early Universe WCU, for hospitality.  
This work has been supported by World Class University grant 
R32-2009-000-10130-0 through the National Research Foundation, Ministry 
of Education, Science and Technology of Korea, and in part by the Director, 
Office of Science, Office of High Energy Physics, of the 
U.S.\ Department of Energy under Contract No.\ DE-AC02-05CH11231.  
JW is acknowledging support from the Trans-Regional Collaborative Research
Center TRR 33 ``The Dark Universe'' of the Deutsche Forschungsgemeinschaft (DFG).

\appendix 

\section{Influence of Perturbations on Matter Growth} \label{sec:apxpert} 

As discussed in the Introduction, for dark energy perturbations to influence 
growth of large scale structure one requires the dark energy equation of 
state to be sufficiently different from $-1$ at a time when there is 
nonnegligible dark energy density, and a low dark energy sound speed.  This 
led to consideration of the class of cold early dark energy models.  If 
one only includes the expansion effects of the dark energy on the matter 
growth, and not the impact of the dark energy perturbations, for example 
through solving the usual second order differential equation for the 
matter density perturbation $\delta_m$ sourced only by itself, then one 
obtains inaccurate results relative to solving the coupled linear Boltzmann 
equations, as in for example CAMB.  

\begin{figure}[htbp!]
\includegraphics[width=\columnwidth]{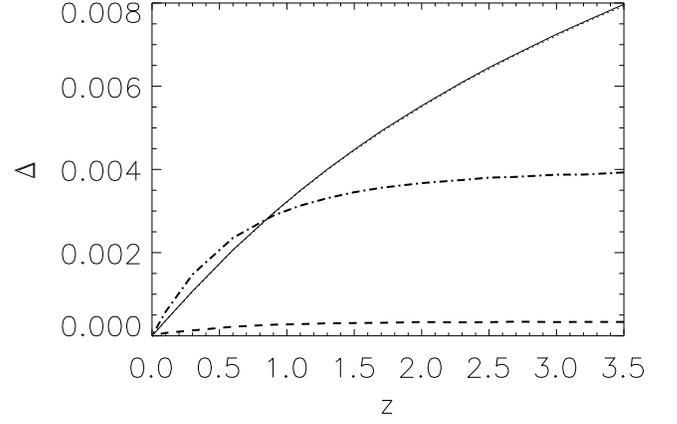}
\caption{Neglect of dark energy perturbations causes deviations in 
estimation of the 
matter growth given by $\Delta=[D(z)/D(0)]/\sqrt{P(z)/P(0)}-1$, 
where the linear matter 
power spectrum $P$ is calculated including the perturbations and the growth 
factor $D$ neglects them.  The case of cold early dark energy with 
$\Oe=0.01$, $c_s=0$, $w_0=-0.99$ is shown by the solid (nearly identical 
dotted) curve for wavenumber $k=0.2\,h$/Mpc ($0.005\,h$/Mpc).  By contrast, 
the no early dark energy $w=-0.99$, $c_s=0$ case (dashed curve) has 
negligible perturbations, and the $w=-0.9$, $c_s=0$ case (dot-dashed curve) 
nearly so. 
}
\label{fig:ratio}
\end{figure}

Here we quantify this deviation in the matter growth from neglecting the 
dark energy perturbations.  Figure~\ref{fig:ratio} shows the growth factor 
$D(z)$ obtained from solving the usual second order differential equation 
for growth, $\ddot D+2H\dot D-4\pi G\rho_m(a) D=0$, relative to the growth 
amplitude of the density power spectrum, $\sqrt{P(z)}$ from 
CAMB modified for cold early dark energy, all normalized to today.  For a 
canonical cold early dark energy case (such as in the main text) the 
inaccuracy is of order $(\Oe/0.02)\%$ at $z=1.5$, scale independent for 
$k\gtrsim0.005\,h$/Mpc.  Dark energy, early or not, with $c_s=1$ 
has negligible perturbations on these scales, and so negligible deviation. 
When there is no early dark energy 
then the deviation is roughly proportional to $(1+w)/(1-3w)$ for constant 
$w$ with $c_s=0$ (see \cite{dhl}), and below $0.5\%$ for $w<-0.9$.

\section{Perturbations When Crossing $w=-1$} \label{sec:apxcross} 

Given the role that dark energy perturbations can play in structure 
growth, it is of interest to ensure the perturbations remain well behaved. 
When $w$ crosses $-1$ (which it does not do for the models considered 
in the main text), it is not obvious from the Boltzmann evolution equations 
for the density and velocity perturbations that good behavior is guaranteed. 
Here we present an analysis demonstrating the conditions under which the 
crossing does not disrupt the perturbation evolution, including the effect 
of sound speed behavior. Detailed discussions of the behaviour of dark energy at the $(1+w) = 0$ barrier
can be found in \cite{Hu:2004kh,Li:2010hm,Fang:2008sn}. 

Adopting the conventions of \cite{mabert} (MB), the metric is 
\be 
g_{00} = -a^{2} ( 1+ 2\psi),\ g_{0i}=0,\ 
g_{ij} = a^{2} (1-2\phi) \delta_{ij} \ , 
\ee 
and the perturbation equations are 
\begin{eqnarray} \label{eq:e1} 
& & {k^{2} \over H_{0}^{2}} \phi + 3 \h^{2} \left( \psi + \phi'  \right) = -{4\pi G a^{2} \over H_{0}^{2}}\sum_{i} \rho_{i} \delta_{i} \\ 
& &  \phi = \psi \\ 
 \label{eq:m1} & &  \h \delta' = -(1+w) \left(\theta - 3\h \phi'  \right) - 3\h \left( {\delta P \over \delta \rho} - w \right) \delta \\ 
\label{eq:m2} & &  \h \theta' = -\h (1-3w) \theta -  \h {w' \over 1+w} \theta + {\delta P/ \delta \rho \over 1+w} {k^{2} \over H_{0}^{2}}\delta +  {k^{2} \over H_{0}^{2}} \psi \\
 & &  \h \delta'_{\rm m} = - \left(\theta_{\rm m} - 3\h \phi'  \right) \\ & &  \h \theta'_{\rm m} = -\h  \theta_{\rm m}  +  {k^{2} \over H_{0}^{2}} \psi \\ 
& & 2\h \h' = -(1+3w) \left( \h^{2} - {\Omega_{\rm m} \over a} - {\Omega_{\rm rad} \over a^{2}} \right) - {\Omega_{\rm m} \over a} - 2{\Omega_{\rm rad} \over a^{2}}  \\ \label{eq:e5} & & \rho'_{\rm i} = -3(1+w_{\rm i}) \rho_{\rm i}  
\end{eqnarray} 
where primes denote differentiation with respect to $N=\ln a$ (or 
equivalently $x$), $\theta = \theta_{MB}/H_{0}$, $c_{\rm s}^{2} = 
\delta P/\delta \rho$, and $\h = {\cal H}/{\cal H}_{0}$. To solve this 
system of equations in the vicinity of the time $N_s$ when the evolving 
dark energy equation crosses $w=-1$, we use a power series expansion  
in $x=N-N_s$ and the method of Frobenius. 

Specifically, we take 
\begin{equation} 
w(x) = -1 + w_{1} x^{\alpha} + {\cal O} \left(x^{\alpha+1} \right) \ . 
\end{equation} 
Generally we expect $\alpha=1$ but we allow the crossing to be at an 
inflection point with integer $\alpha>1$.  We must also consider the 
second free function $c_{\rm s}^{2}$, which we parametrize near the 
crossing as 
\begin{equation} 
c_{\rm s}^{2}(|x|\ll1) = c_{0} x^{\beta} \ , 
\end{equation} 
where $c_{0}$ is a constant and $\beta=0$, $<0$, $>0$ give three 
distinct cases. We take the ansatz 
\begin{eqnarray}  \label{eq:s1} 
& & \delta = \sum_{n=0}^{\infty} \delta_{n} x^{n + k_{\delta}} \qquad 
\theta = \sum_{n=0}^{\infty} \theta_{n} x^{n+k_{\theta}} \\  
& & \delta_{\rm m} = \sum_{n=0}^{\infty} \delta_{\rm m, n} x^{n + k_{\rm m}} 
\quad \theta_{\rm m} = \sum_{n=0}^{\infty} \theta_{\rm m, n} x^{n+k_{\rm m}}\\ 
& & \phi = \sum_{n=0}^{\infty} \phi_{n} x^{n + k_{\phi}}   \label{eq:s3} 
\end{eqnarray} 
where $k_{\delta}, k_{\theta}, k_{\phi}$ are real numbers, determined using 
the indicial equations of the corresponding series solution. We insert 
Eqs.~(\ref{eq:s1}-\ref{eq:s3}) into Eqs.~(\ref{eq:e1}-\ref{eq:e5}).  

Before doing so, we calculate $\rho_{\rm i}$ and $\h$ to second order in $x$ using 
\begin{eqnarray} 
& & \h^{2} = {8\pi G a^{2} \over 3{ H}_{0}^{2}} \left( \rho_{\rm m} + \rho_{\rm rad} + \rho_{\rm de}\right) \\
& & \rho'_{\rm m} = -3 \rho_{\rm m} \\ 
& & \rho'_{\rm rad} = -4 \rho_{\rm rad} \\ 
& & \rho'_{\rm de} = -3 \left( 1+ w(N)\right) \rho_{\rm de} \ . 
\end{eqnarray} 
The solution is given by 
\begin{eqnarray}  
& & \rho_{\rm m} = \rho_{\rm m, s} \left( 1 - 3x + 9 x^{2}+ {\cal O}(x^{3})\right) \\
& & \rho_{\rm rad} = \rho_{\rm r, s} \left( 1 - 4 x + 16 x^{2} + {\cal O}(x^{3})\right) \\
& & \rho_{\rm de} = \rho_{\rm de, s} \left( 1 - 3{w_{1}\over 1+\alpha}x^{\alpha+1}  + {\cal O}(x^{\alpha+2})\right) \\ 
& & \h = \left[{8\pi G e^{2N_{\rm s}} (\rho_{\rm m,s}+\rho_{\rm r,s}+\rho_{\rm de,s}) \over 3 {H}_{0}^{2}}\right]^{1/2} \\ 
& & \quad \times \left[ 1 - \left( {3\rho_{\rm m,s} + 4 \rho_{\rm r,s}  \over 2\rho_{\rm m,s} + 2 \rho_{\rm r,s} + 2\rho_{\rm de,s}} - {1 \over 2N_{\rm s}}\right)x \right]  + {\cal O}(x^{2}) \notag 
\end{eqnarray}

We can now substitute the ansatz for $\delta, \theta$ and the above expansions into the perturbation equations.  Assuming the matter perturbations are 
well behaved during the dark energy crossing implies $k_{\rm m} = 0$ and 
$k_{\phi}=0$. Hence $\delta_{\rm m}$, $\theta_{\rm m}$ and $\phi$ all 
approach constant values at the crossing $x=0$. 

To solve for the remaining variables, we use the expansions for $\theta$ and $\delta$ in Eq.~($\ref{eq:m1}$). Our approach will be to combine the equations for $\theta$ and $\delta$ by differentiating Eq.~($\ref{eq:m2}$) and 
then removing $\delta$ and $\delta'$. The resulting second order equation for $\theta$ will yield the indicial equation, the solutions of which will correspond to the leading order behaviour of $\theta$. This can then be used in 
Eq.~($\ref{eq:s1}$) to obtain $\delta$. Keeping only the most singular 
terms, we find the following equation for $k_{\theta}$,  
\begin{equation} 
k_{\theta}^{2}+(2\alpha-\beta -1)k_{\theta} - \alpha(\beta-\alpha+1) = 0 \ , 
\end{equation} 
and $k_{\delta} = 0$.  This implies the dark energy density perturbation 
stays constant in an infinitesimal interval around the crossing, but the 
momentum perturbation may diverge.  The two roots of the equation are 
\be 
k_\theta=1+\beta-\alpha \quad ; \quad k_\theta=-\alpha 
\ee 
but the second root always leads to divergence.  The first root gives 
stable perturbations for $\beta\ge\alpha-1$.  If $\beta<0$, i.e.\ the 
sound speed 
diverges at the crossing, then the momentum does as well for all $\alpha$.  

The key criterion 
\be 
\beta\ge\alpha-1 
\ee 
can be seen heuristically from the $\theta'$ equation.  Terms on the 
right hand side can diverge no more severely than $x^{-1}$ in order 
that the integration over $x$ to get $\theta$ gives a bounded $\theta$. 
For the term involving $w'/(1+w)\sim x^{-1}$ this is fulfilled for all 
$\alpha$, but the sound speed term gives $x^\beta/x^\alpha=x^{\beta-\alpha}$ 
so we require $\beta-\alpha\ge -1$, precisely the criterion above. 

The next to leading order behaviour is dependent upon the values of $\alpha,\beta$. If we concentrate 
on the case $(\alpha,\beta) = (1,0)$, then we find 
\begin{eqnarray} 
& & \delta_{1} = -\frac{w_{1}}{\h_{\rm s}} \theta_{0}-3(1+c_{0})\delta_{0} \\
 & & \theta_{1} = -4 \theta_{0} + \frac{c_{0}}{w_{1} \h_{\rm s}} 
\frac{k^{2}}{{H}_{0}^{2}} \delta_{0}  \ . 
\end{eqnarray}  
The zeroth order coefficients $\delta_0$ and $\theta_0$ are determined 
by the two initial conditions required for the two first order equations. 
The divergent second root $\theta\sim x^{-\alpha}$ can be removed by setting 
the initial condition such that $\theta_0=0$.  The dark energy perturbations 
in the vicinity of the crossing are therefore well behaved for this case, 
with 
\begin{eqnarray} 
& & \delta=\delta_{0} \left( 1 - 3(1+c_{0}) x + {\cal O} (x^{2}) \right) \\ 
& & \theta = \delta_{0} \left( \frac{c_{0}}{w_{1} \h_{\rm s}} 
\frac{k^{2}}{{ H}_{0}^{2}} + {\cal O} (x) \right) \ . 
\end{eqnarray}  

\noindent Of course, in numerical studies it is not possible to evade the divergent root of $\theta$ by choosing initial conditions
appropriately, and so one must resort to other means 
\cite{Hu:2004kh,Li:2010hm}. The underlying physics of the singular 
solution at the crossing is well understood \cite{Fang:2008sn}.



\begin{thebibliography}{99}

\bibitem{suzuki} 
N. Suzuki et al., ApJ 746, 85 (2012) [arXiv:1105.3470] 

\bibitem{wigglez} 
C. Blake et al., MNRAS 425, 405 (2012) [arXiv:1204.3674] 

\bibitem{boss} 
B.A. Reid et al., MNRAS 426, 2719 (2012) [arXiv:1203.6641] 

\bibitem{2003MNRAS.346..987W}
J. Weller, A. Lewis, MNRAS, 346, 987 (2003)

\bibitem{2004PhRvD..69h3503B}
R. Bean, O. Dor\'e, PRD, 69, 083503 (2004)

\bibitem{09010916} 
R. de Putter, O. Zahn, E.V. Linder, Phys. Rev. D 79, 065033 (2009) 
[arXiv:0901.0916] 

\bibitem{dhl} 
R. de Putter, D. Huterer, E.V. Linder, Phys. Rev. D 81, 103513 (2010) 
[arXiv:1002.1311] 

\bibitem{10105612} 
E. Calabrese, R. de Putter, D. Huterer, E.V. Linder, A. Melchiorri, 
Phys. Rev. D 83, 023011 (2011) [arXiv:1010.5612] 

\bibitem{11034132} 
E. Calabrese, D. Huterer, E.V. Linder, A. Melchiorri, L. Pagano, 
Phys. Rev. D 83, 123504 (2011) [arXiv:1103.4132] 

\bibitem{11060299} 
S. Joudaki \& M. Kaplinghat, Phys. Rev. D 86, 023526 (2012) [arXiv:1106.0299] 

\bibitem{11105328} 
C.L. Reichardt, R. de Putter, O. Zahn, Z. Hou, ApJ 749, L9 (2012) 
[arXiv:1110.5328] 

\bibitem{Wetterich04}
C. Wetterich, Phys. Lett. B 594, 17 (2004) [arXiv:astro-ph/0403289] 

\bibitem{doranrob}
M. Doran \& G. Robbers, JCAP 0606, 026 (2006) [arXiv:astro-ph/0601544] 

\bibitem{fll1} 
M.J. Francis, G.F. Lewis, E.V. Linder, MNRAS 394, 605 (2008) 
[arXiv:0808.2840] 

\bibitem{grossi} 
M. Grossi \& V. Springel, MNRAS 394, 1559 (2009) [arXiv:0809.3404] 

\bibitem{fll2} 
M.J. Francis, G.F. Lewis, E.V. Linder, MNRAS Letters 393, L31 (2008) 
[arXiv:0810.0039] 

\bibitem{alam} 
U. Alam, Z. Luki{\'c}, S. Bhattacharya, ApJ 727, 87 (2011) [arXiv:1004.0437] 

\bibitem{10072188} 
D. Sapone, M. Kunz, L. Amendola, Phys. Rev. D 82, 103535 (2010) 
[arXiv:1007.2188] 

\bibitem{linjen} 
E.V. Linder \& A. Jenkins, MNRAS 346, 573 (2003) [arXiv:astro-ph/0305286] 

\bibitem{10045509} 
G. Ballesteros \& J. Lesgourgues, JCAP 1010, 014 (2010) [arXiv:1004.5509] 

\bibitem{Oyaizu:08}
H. Oyaizu, Phys. Rev. D 78, 123523 (2008) [arXiv:0807.2449] 
	
\bibitem{Li:12}
B. Li, G.-B. Zhao, R. Teyssier, K. Koyama, JCAP 01, 051 (2012) 
[arXiv:1110.1379] 

\bibitem{Puchwein:13}
E. Puchwein, M. Baldi, V. Springel, arXiv:1305.2418 

\bibitem{tinker} 
J.L. Tinker, A.V. Kravtsov, A. Klypin, K. Abazajian, M.S. Warren, G. Yepes, 
S. Gottlober, D.E. Holz, ApJ 688, 709 (2008) [arXiv:0803.2706] 

\bibitem{camb} 
A. Lewis, A. Challinor, \& A. Lasenby, ApJ 538, 473 (2000) 
[arXiv:astro-ph/9911177]; \url{http://camb.info} 

\bibitem{jenkins} 
A. Jenkins, C.S. Frenk, S.D.M. White, J.M. Colberg, S. Cole, A.E. Evrard, 
H.M.P. Couchman, N. Yoshida, MNRAS 321, 372 (2001) [arXiv:astro-ph/0005260] 

\bibitem{shethtormen} 
R.K. Sheth \& G. Tormen, MNRAS 308, 119 (1999) [arXiv:astro-ph/9901122] 

\bibitem{warren} 
M.S. Warren, K. Abazajian, D.E. Holz, L. Teodoro, ApJ 646, 881 (2006) 
[arXiv:astro-ph/0506395] 

\bibitem{pace} 
  R.~C.~Batista and F.~Pace,
  arXiv:1303.0414 [astro-ph.CO].

\bibitem{linderrob} 
E.V. Linder \& G. Robbers, JCAP 0806, 004 (2008) [arXiv:0803.2877] 

\bibitem{euclid} 
R. Laureijs et al, ESA/SRE(2011)12 [arXiv:1110.3193]; 
\url{http://www.euclid-ec.org} 

\bibitem{des} 
\url{http://www.darkenergysurvey.org}

\bibitem{Cunha:2009dz}
  C.~Cunha, D.~Huterer and J.~A.~Frieman,
  Phys.\ Rev.\ D {\bf 80} (2009) 063532
  [arXiv:0904.1589 [astro-ph.CO]].
  

\bibitem{Cunha:2009rx}
  C.~E.~Cunha and A.~E.~Evrard,
  Phys.\ Rev.\ D {\bf 81} (2010) 083509
  [arXiv:0908.0526 [astro-ph.CO]].

\bibitem{Lima:2005tt}
  M.~Lima and W.~Hu,
  Phys.\ Rev.\ D {\bf 72} (2005) 043006
  [astro-ph/0503363].
  
\bibitem{Rykoff:2011xi}
  E.~S.~Rykoff, B.~P.~Koester, E.~Rozo, J.~Annis, A.~E.~Evrard, S.~M.~Hansen, J.~Hao and D.~E.~Johnston {\it et al.},
  Astrophys.\ J.\  {\bf 746} (2012) 178
  [arXiv:1104.2089 [astro-ph.CO]].

\bibitem{Biviano}
A. Biviano, private communication and for the Euclid Red Book

\bibitem{Rozo}
E. Rozo, private communication.

\bibitem{Ade:2013zuv}
  P.~A.~R.~Ade {\it et al.}  [Planck Collaboration],
  arXiv:1303.5076 [astro-ph.CO].

\bibitem{planck} 
P.A.R. Ade et al, A\&A, 536, A1 (2011); 
\url{http://planck.esa.int} 

\bibitem{13051942} 
D. Sapone, E. Majerotto, M. Kunz, B. Garilli, arXiv:1305.1942 

\bibitem{wmap9} 
C.L. Bennett et al, arXiv:1212.5225 

\bibitem{cosmomc} 
A. Lewis \& S. Bridle, Phys. Rev. D 66, 103511 (2002) 
[arXiv:astro-ph/0205436] \\ 
\url{http://cosmologist.info/cosmomc}

\bibitem{vallinotto} 
A. Vallinotto, arXiv:1304.3474 

\bibitem{Hu:2004kh}
  W.~Hu,
  Phys.\ Rev.\ D {\bf 71} (2005) 047301
  [astro-ph/0410680].
  
\bibitem{Li:2010hm}
  M.~Li, Y.~Cai, H.~Li, R.~Brandenberger and X.~Zhang,
  Phys.\ Lett.\ B {\bf 702} (2011) 5
  [arXiv:1008.1684 [astro-ph.CO]].
  
\bibitem{Fang:2008sn}
  W.~Fang, W.~Hu and A.~Lewis,
  Phys.\ Rev.\ D {\bf 78} (2008) 087303
  [arXiv:0808.3125 [astro-ph]].
  
\bibitem{mabert} 
C-P. Ma \& E. Bertschinger, ApJ 455, 7 (1995) [arXiv:astro-ph/9506072] 


\end{thebibliography}
\end{document}